\documentclass{epl2}
\usepackage{amsfonts}
\usepackage{amssymb}

\title{Determination of the critical exponents for the isotropic-nematic phase transition in a system of long rods on
two-dimensional lattices: universality of the transition}
\shorttitle{Long rods on 2-D lattices}

\author{D. A. Matoz-Fernandez \and D. H. Linares \and A. J. Ramirez-Pastor}

\institute{
   Departamento de F\'{\i}sica, Instituto de F\'{\i}sica Aplicada, Universidad Nacional de San
Luis-CONICET, Chacabuco 917, 5700 San Luis, Argentina
 }


\abstract{ Monte Carlo simulations and finite-size scaling
analysis have been carried out to study the critical behavior and
universality for the isotropic-nematic phase transition in a
system of long straight rigid rods of length $k$ ($k$-mers) on
two-dimensional lattices. The nematic phase, characterized by a
big domain of parallel $k$-mers, is separated from the isotropic
state by a continuous transition occurring at a finite density.
The determination of the critical exponents, along with the
behavior of Binder cumulants, indicate that the transition belongs
to the 2D Ising universality class for square lattices and the
three-state Potts universality class for triangular lattices.}

\begin{document}

\maketitle

\section{Introduction}

The study of systems of large particles in solution is one of the
central problems in statistical mechanics and has been attracting
a great deal of interest since long ago. Onsager~\cite{ONSAGER}
predicted that very long and thin rods interacting with only
excluded volume interaction can lead to long-range orientational
(nematic) order. The nematic phase, characterized by a big domain
of parallel molecules, is separated from an isotropic state by a
phase transition occurring at a finite critical density. Computer
simulations of hard ellipses of finite length~\cite{BARON}
confirmed the Onsager's classic prediction that particle shape
anisotropy can be a sufficient condition to induce the long-range
orientational order found in nematic liquid crystals.

Flory~\cite{FLORY} and Huggins~\cite{HUGGI} studied a system of
long rod-like molecules by means of a lattice calculation. The
approach, which is a direct generalization of the theory of binary
liquids in two dimensions, indicates that the lattice model would
also show an isotropic-nematic (I-N) phase transition as a
function of density.

A notable feature is that nematic order is only stable for
sufficiently large aspect ratios while isotropic systems of short
rods not show nematic order at all. The long-range orientational
order also disappears in the case of irreversible adsorption (no
desorption)~\cite{HINRICHSEN}, where the distribution of adsorbed
objects is different from that obtained at
equilibrium~\cite{TALBOT,KONDRAT}. Thus, at high coverage, the
equilibrium state corresponds to a nematic phase with long-range
correlations, whereas the final state generated by irreversible
adsorption is a disordered state (known as jamming state), in
which no more objects can be deposited due to absence of free
space of appropriate size and shape (the jamming state has
infinite memory of the process and orientational order is purely
local). Another important factor in the phase stability of long
rods is the flexibility of the adparticles. This property plays an
important role in systems of, e.g., stiff polymers and linear
micelles and its generic effect is a significant depression of
nematic order compared to rigid particles~\cite{VROEGE}.

For the continuum problem, there is general agreement that in
three dimensions, infinitely thin rods undergo a first-order I-N
transition, as was pointed out by Onsager~\cite{ONSAGER}. In two
dimensions, the nature of the I-N transition depends crucially on
the particle interactions and a rich variety of behaviors is
observed~\cite{STRALEY,FRENKEL}.

In the case of lattice models of straight hard rods of length $k$,
which is the topic of this paper, the inherent complexity of the
system still represents a major difficulty to the development of
approximate solutions, and computer simulations appear as a very
important tool for investigating this subject. In this sense, a
system of straight rigid rods of length $k$ on a square lattice,
with two allowed orientations, was recently studied by Monte Carlo
simulations~\cite{GHOSH}. Ghosh and Dhar found strong numerical
evidence that the system shows nematic order at intermediate
densities for $k \geq 7$. However, the authors were not able to
determine the critical quantities (critical point and critical
exponents) characterizing the I-N phase transition occurring in
the system.

Despite these recent results there is a remaining question to be
answered: ``What type of phase transition is it?"\footnote{Even
though it is theoretically expected that the I-N transition in a
system of long rigid rods on a square (triangular) lattice belongs
to the 2D Ising (three-state Potts) universality
class~\cite{GHOSH}, numerical verification of this has not been
possible so far.} The objective of this Letter is to provide a
thorough study in this direction. For this purpose, extensive
Monte Carlo (MC) simulations supplemented by analysis using
finite-size scaling (FSS) theory~\cite{FISHER,BINDER,PRIVMAN} have
been carried out to study the critical behavior in a system of
long rigid rods deposited on square and triangular lattices. In
the case of FSS analysis, the conventional normalized scaling
variable $\epsilon \equiv T/T_c - 1$ was replaced by $\epsilon
\equiv \theta/\theta_c - 1$, where $T$, $T_c$, $\theta$ and
$\theta_c$ represent temperature, critical temperature, density
and critical density, respectively. A nematic phase, characterized
by a big domain of parallel $k$-mers, is separated from the
disordered state by a continuous phase transition occurring at a
finite critical temperature. Based on the strong axial anisotropy
of the nematic phase, an order parameter measuring the orientation
of the particles has been introduced. Taking advantage of its
definition, we were able to study for the first time the
universality class of the I-N phase transition occurring in the
system. The accurate determination of the critical exponents,
along with the behavior
 of Binder cumulants, confirmed that the transition of rigid rods on
square (triangular) lattices, with two (three) allowed
orientations, belongs to the 2D Ising (three-state Potts)
universality class.

\section{Model and Monte Carlo method}

As in Ref.~\cite{GHOSH}, we address the general case of adsorbates
assumed to be linear rigid particles containing $k$ identical
units ($k$-mers), with each one occupying a lattice site. Small
adsorbates would correspond to the monomer limit ($k = 1$). The
distance between $k$-mer units is assumed to be equal to the
lattice constant; hence exactly $k$ sites are occupied by a
$k$-mer when adsorbed. The only interaction between different rods
is hard-core exclusion: no site can be occupied by more than one
$k$-mer. The surface is represented as an array of $M = L \times
L$ adsorptive sites in a square or triangular lattice arrangement,
where $L$ denotes the linear size of the array.

The degree of order in the adsorbed phase is calculated for each
configuration according to the standard method used for the Potts
model~\cite{WU}. To this end, we first build a set of vectors
$\{\vec{v}_1,\vec{v}_2,\cdots,\vec{v}_n\}$ with the following
properties: $(i)$ each vector is associated to one of the $n$
possible orientations (or directions) for a $k$-mer on the
lattice; $(ii)$ the $\vec{v}_i$'s lie in a two-dimensional space
(or are co-planar) and point radially outward from a given point
$P$ which is defined as coordinate origin; $(iii)$ the angle
between two consecutive vectors, $\vec{v}_i$ and $\vec{v}_{i+1}$,
is equal to $2\pi/n$; and $(iv)$ the magnitude of $\vec{v}_i$ is
equal to the number of $k$-mers aligned along the $i$-direction.
Note that the $\vec{v}_i$'s have the same directions as the $q$
vectors in Ref.~\cite{WU}. These directions are not coincident
with the allowed directions for the $k$-mers on the real lattice.
Then the order parameter $\delta$ of the system is given by
\begin{equation}
\delta =  \frac{\left | \sum_{i=1}^n \vec{v}_i \right
|}{\sum_{i=1}^n \left | \vec{v}_i \right |}
 \label{fi}
\end{equation}
$\delta$ represents a general order parameter measuring the
orientation of the $k$-mers on a lattice with $n$ directions. In
the case of square lattices, $n=2$ and the angle between
$\vec{v}_1$ and $\vec{v}_{2}$ is $\pi$ [see Fig. 1(a)].
Accordingly, the order parameter reduces to $\delta =  \left | v_1
-v_2 \right |/ \left ( v_1+v_2 \right )$, $v_1$ ($v_2$) being the
number of $k$-mers aligned along the horizontal (vertical)
direction. This expression coincides with the order parameter $Q$
defined in Ref.~\cite{GHOSH}. On the other hand, $n=3$ for
triangular lattices and $\vec{v}_1$, $\vec{v}_{2}$ and
$\vec{v}_{3}$ form angles of $2\pi/3$ between them [see Fig.
1(b)].

\begin{figure}
\includegraphics[width=6.5cm,clip=true]{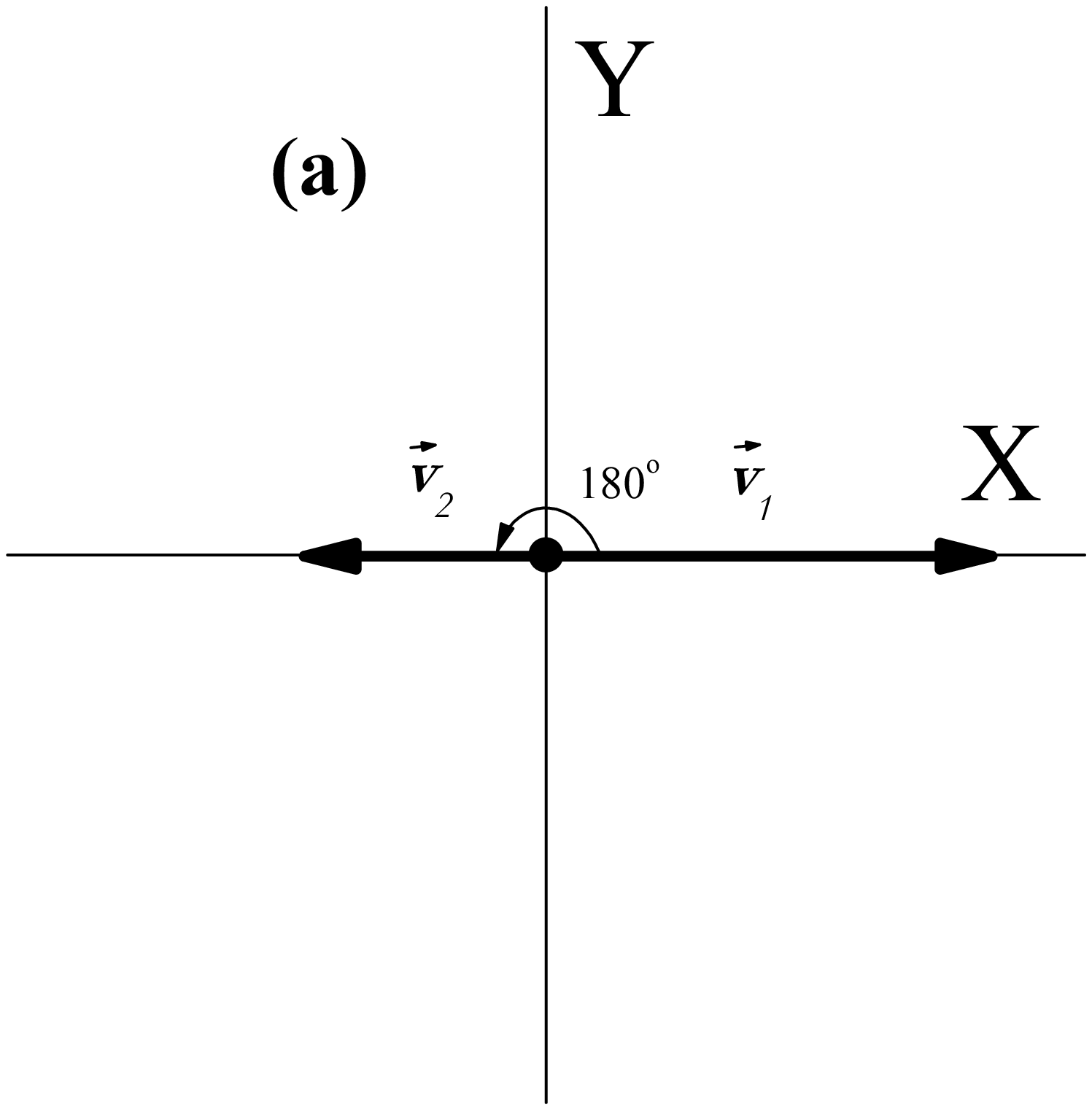}
\includegraphics[width=6.5cm,clip=true]{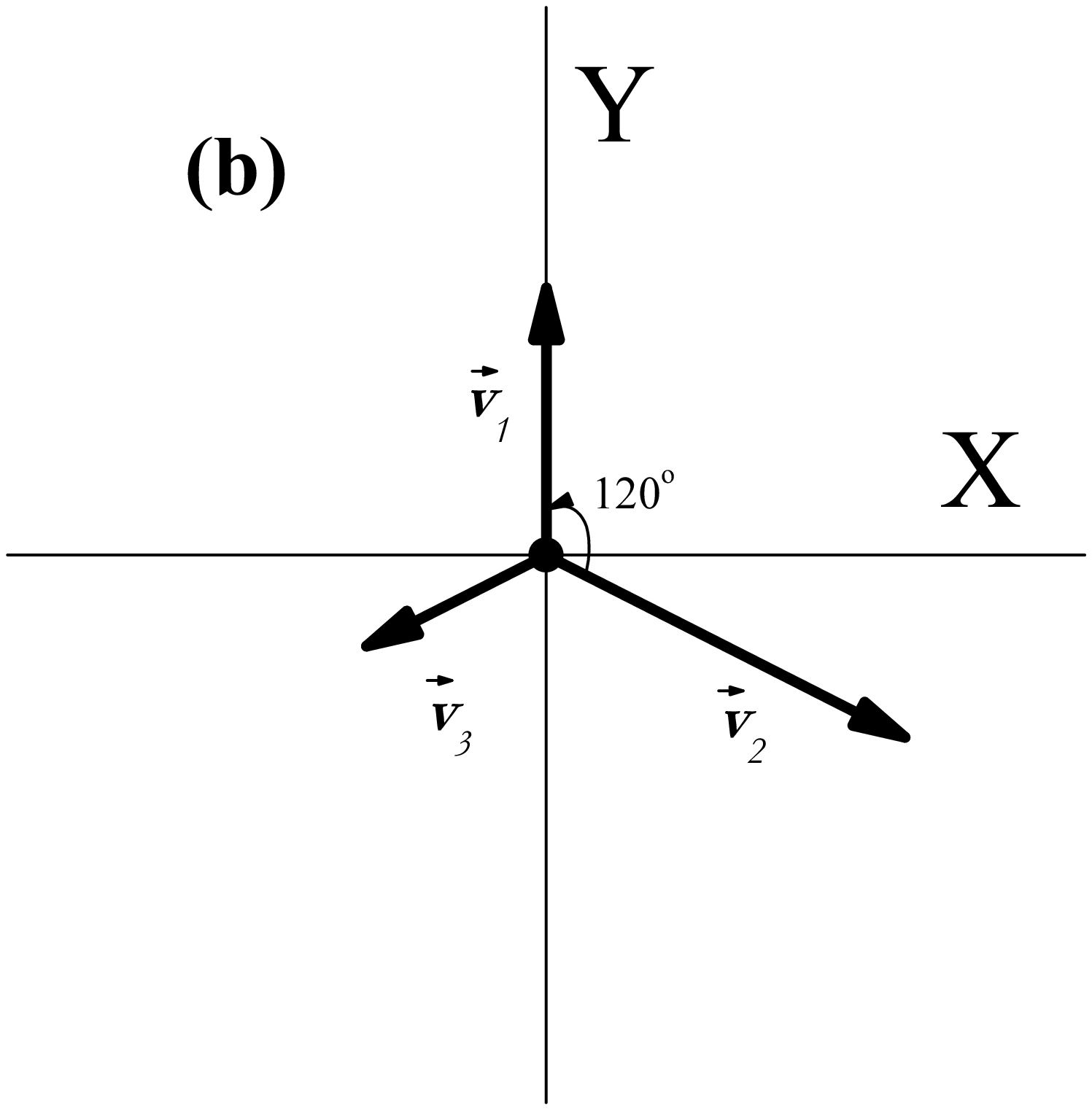}
\caption{\label{figure0} Schematic representation of the set of
vectors $\{\vec{v}_1,\vec{v}_2,\cdots,\vec{v}_n\}$ for square (a)
and triangular (b) lattices.}
\end{figure}

When the system is disordered $(\theta<\theta_c)$, all
orientations are equivalents and $\delta$ is zero. As the density
is increased above $\theta_c$, the $k$-mers align along one
direction and $\delta$ is different from zero. Thus, $\delta$
appears as a proper order parameter to elucidate the phase
transition.

The problem has been studied by grand canonical Monte Carlo
simulations using an adsorption-desorption algorithm. The
procedure is as follows. At each time-step, a linear $k$-uple of
nearest-neighbor sites is chosen at random. Then, if the $k$ sites
are empty, an attempt is made to deposit a rod with probability
$p$; if the $k$ sites are occupied by units belonging to the same
$k$-mer, an attempt is made to desorb this $k$-mer with
probability $1-p$; and otherwise, the attempt is rejected. A Monte
Carlo step (MCS) is achieved when $M$ $k$-uples of sites have been
tested to change its occupancy state. Typically, the equilibrium
state can be well reproduced after discarding the first $10^6$
MCS. Then, the next $3 \times 10^6$ MCS are used to compute
averages.


In our Monte Carlo simulations, we varied the adsorption
probability $p$ and monitored the density $\theta$ and the order
parameter $\delta$, which can be calculated as simple averages.
The quantities related with the order parameter, such as the
susceptibility $\chi$, and the reduced fourth-order cumulant $U_L$
introduced by Binder~\cite{BINDER} were calculated as:
\begin{equation}
\chi = \frac{L^2}{k_BT} [ \langle \delta^2 \rangle - \langle
\delta \rangle^2] \label{chi}
\end{equation}
and
\begin{equation}
U_L = 1 -\frac{\langle \delta^4\rangle} {3\langle
\delta^2\rangle^2}, \label{ul}
\end{equation}
where $\langle \cdots \rangle$ means the average over the MC
simulation runs. All calculations were carried out using the BACO
parallel cluster (composed by  60 PCs each with a 3.0 GHz
Pentium-4 processor) located  at Laboratorio de Ciencias de
Superficies y Medios Porosos, Universidad Nacional de San Luis,
San Luis, Argentina. The total CPU time for the present study is
estimated at about 900 days on one node of the BACO cluster.

\begin{figure}
\includegraphics[width=7.5cm,clip=true]{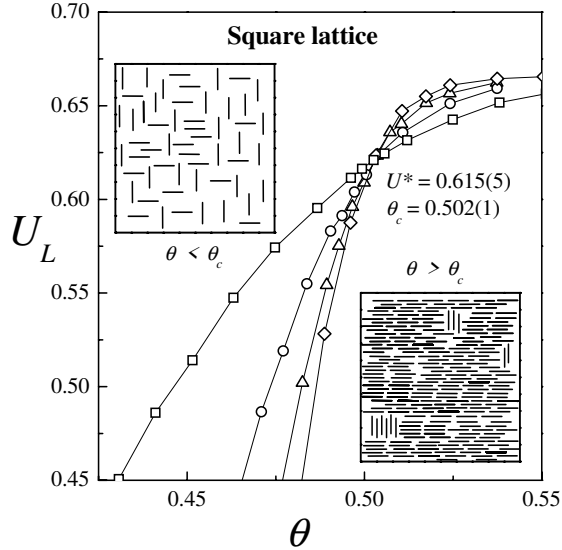}
\caption{\label{figure1} Curves of $U_L(\theta)$ vs $\theta$ for
square lattices of different sizes: squares, L = 50; circles, L =
100; triangles, L = 150 and diamonds, L = 200. From their
intersections one obtains $\theta_c$. Inset in the upper-left
(lower-right) corner shows a typical configuration in the
isotropic (nematic) phase.}
\end{figure}

\section{Results}

\begin{figure}
\includegraphics[width=6.5cm,clip=true]{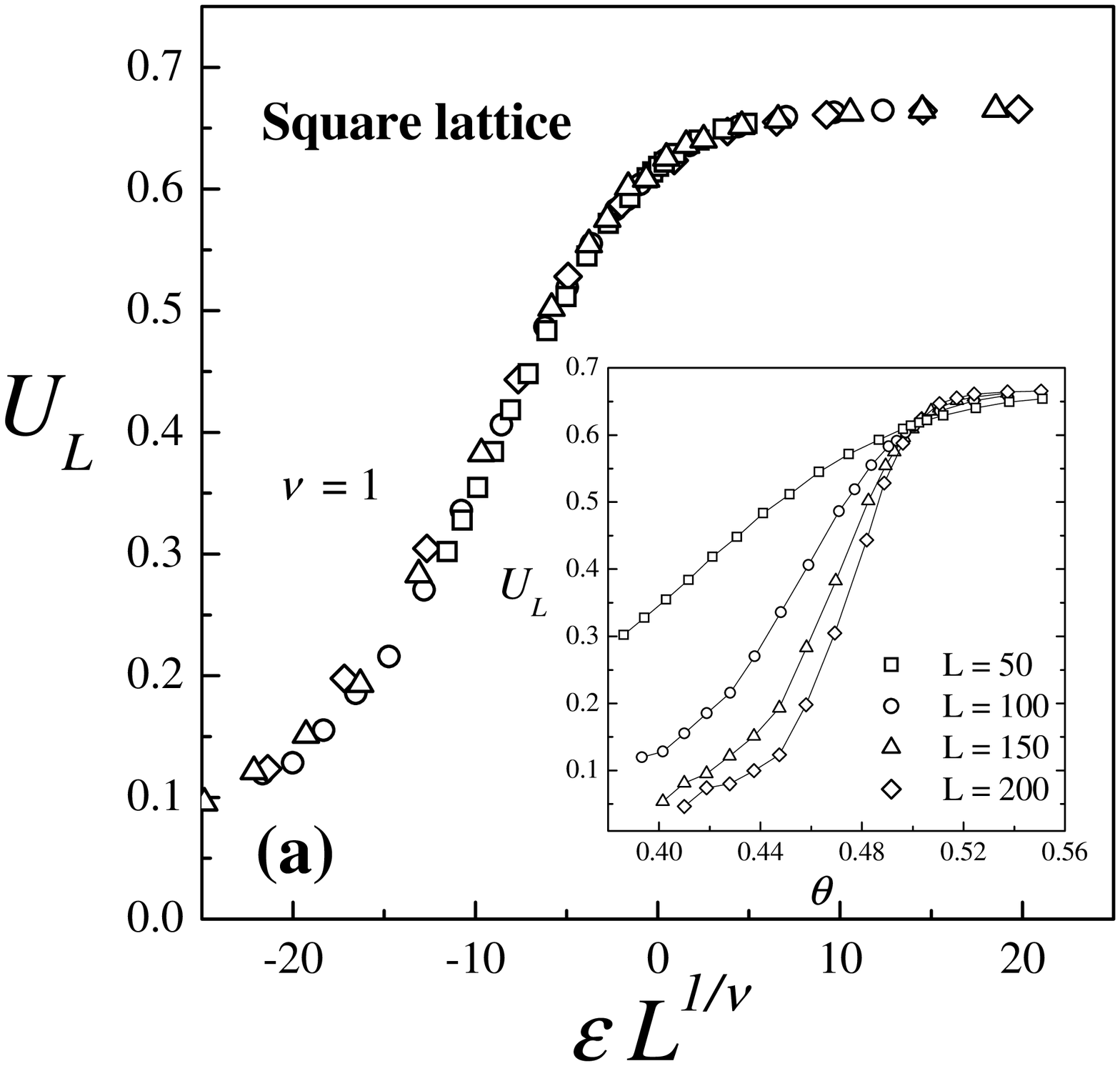}
\includegraphics[width=6.5cm,clip=true]{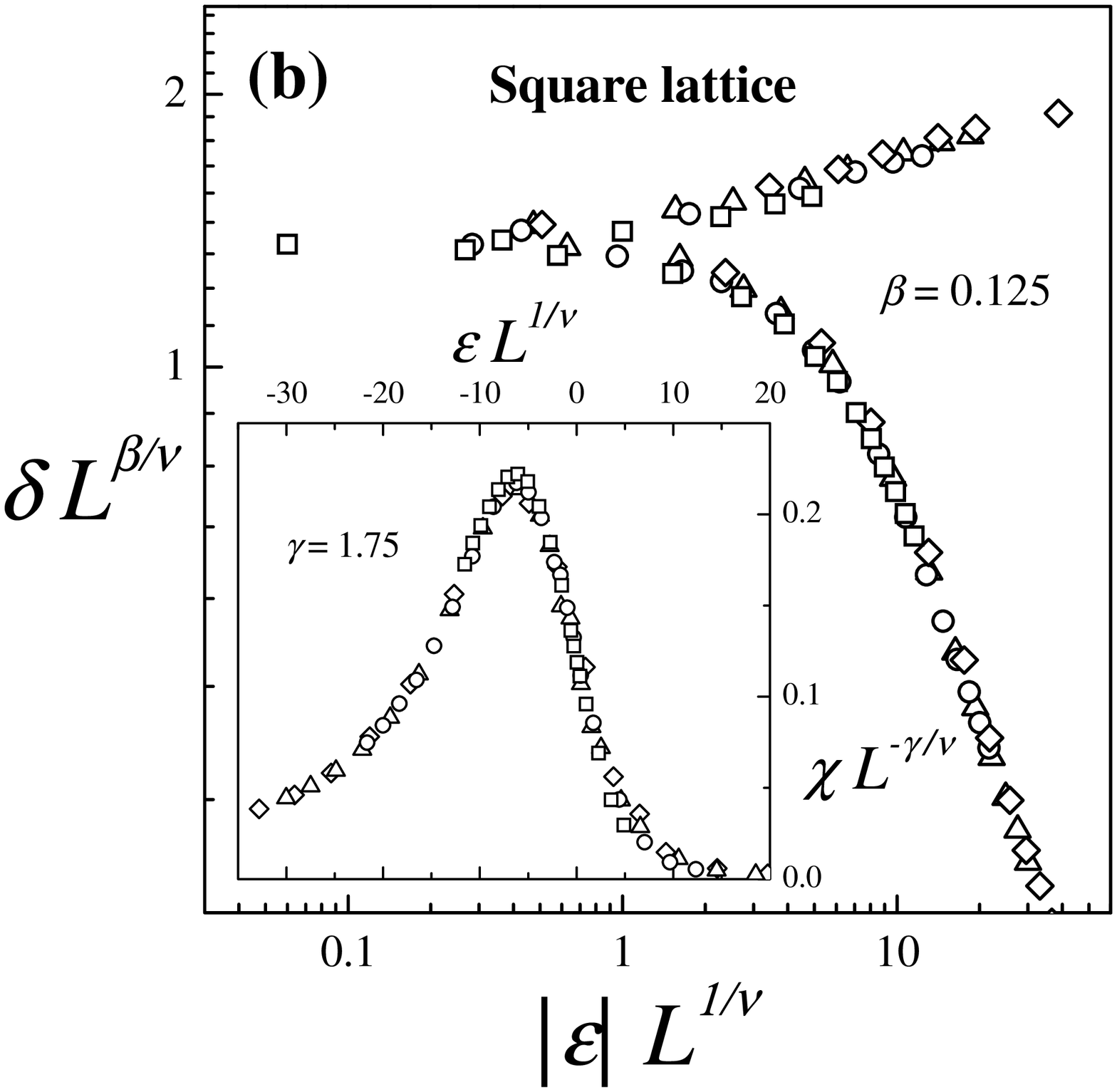}
\caption{\label{figure2} (a) Data collapsing, $U_L$ vs $\epsilon
L^{1/\nu}$, for the cumulants in Fig.~\ref{figure1}. In the inset,
the data in Fig.~\ref{figure1} are plotted over a wide range of
coverage. (b) Data collapsing of the order parameter, $\delta
L^{\beta/\nu}$ vs $|\epsilon| L^{1/\nu}$, and of the
susceptibility, $\chi L^{-\gamma/\nu}$ vs $\epsilon L^{1/\nu}$
(inset). The plots were made using $\theta_c=0.502$ and the exact
2D Ising exponents $\nu=1$, $\beta=0.125$ and $\gamma=1.75$.}
\end{figure}

The critical behavior of the present model has been investigated
by means of the computational scheme described in the previous
paragraphs and finite-size scaling analysis. The FSS theory
implies the following behavior of $\delta$, $\chi$ and $U_L$ at
criticality:
\begin{equation}
\delta = L^{-\beta/\nu} \tilde \delta(L^{1/\nu} \epsilon),
\label{ds}
\end{equation}
\begin{equation}
\chi= L^{\gamma/\nu}\tilde \chi(L^{1/\nu} \epsilon) \label{chis}
\end{equation}
and
\begin{equation}
U_L=\tilde U_L(L^{1/\nu} \epsilon), \label{uls}
\end{equation}
for $L \rightarrow \infty$, $\epsilon \rightarrow 0$ such that
$L^{1/\nu} \epsilon $= finite. Here $\beta$, $\gamma$ and $\nu$
are the standard critical exponents of the order parameter
($\delta \sim -\epsilon^{\beta} $ for $\epsilon\rightarrow 0^-$,
$L\rightarrow \infty$),
 susceptibility($\chi \sim |\epsilon|^\gamma$ for $\epsilon \rightarrow 0$, $L\rightarrow \infty$) and
correlation length $\xi$ ($\xi \sim |\epsilon|^{-\nu}$ for
$\epsilon \rightarrow 0, L \rightarrow \infty$), respectively.
$\tilde \delta, \tilde \chi $ and $\tilde U_L$ are scaling
functions for the respective quantities.

In the case of conventional FSS theory, when the phase transition
is temperature driven, $\epsilon \equiv T/T_c - 1$. In our study,
we modified the standard FSS analysis by replacing temperature by
density. Under this condition, $\epsilon \equiv \theta/\theta_c -
1$.

The calculations were developed for linear $10$-mers ($k=10$).
With this value of $k$, it is expected the existence of a nematic
phase at intermediate densities. The surface was represented as an
array of adsorptive sites in a square or triangular lattice
arrangement. In addition, conventional periodic boundary
conditions were considered. The effect of finite size was
investigated by examining square lattices with $L=50, 100, 150,
200$ and triangular lattices with $L=100, 150, 200, 250$, with an
effort reaching almost the limits of our computational
capabilities.

\begin{figure}
\includegraphics[width=6.5cm,clip=true]{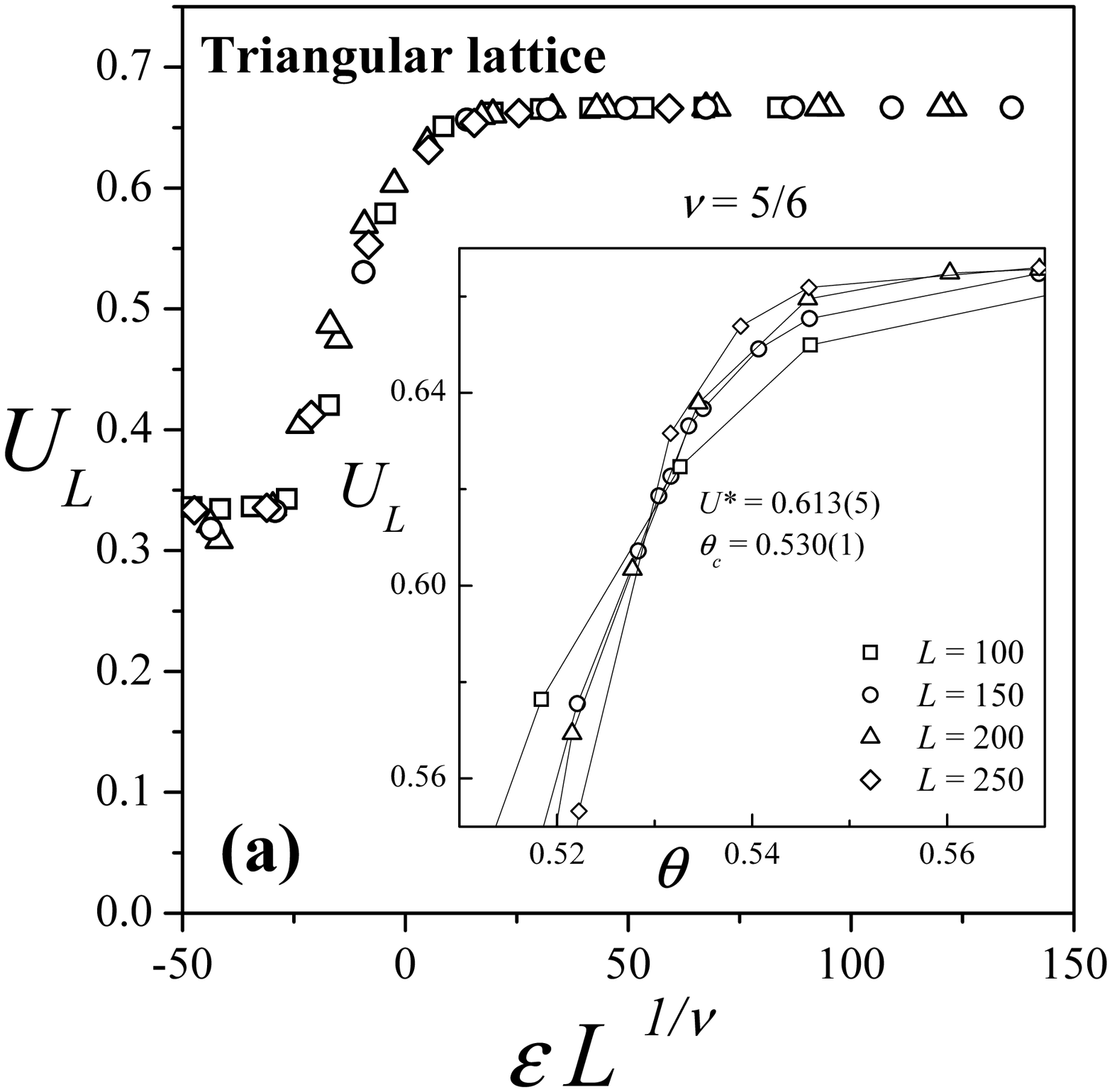}
\includegraphics[width=6.5cm,clip=true]{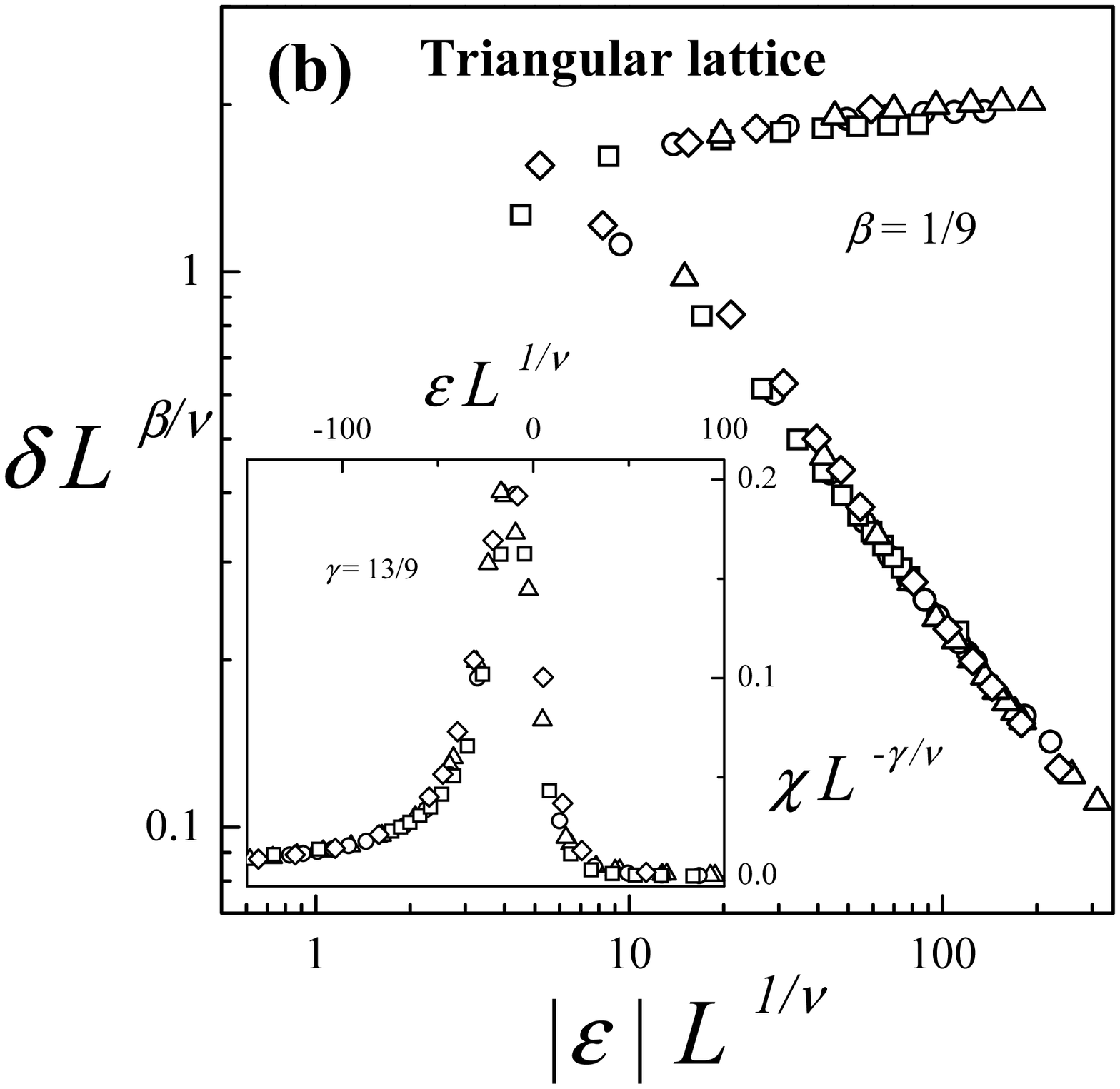}
\caption{\label{figure3} (a) Data collapsing of the cumulants for
triangular lattices. The corresponding curves of $U_L(\theta)$ vs
$\theta$ are shown in the inset. (b) Same as Fig.~\ref{figure2}(b)
for triangular lattices. The plots were made using
$\theta_c=0.530$ and the exact three-state Potts model exponents
$\nu=5/6$, $\beta=1/9$ and $\gamma=13/9$.}
\end{figure}

The critical density has been estimated from the plots of the
reduced four-order cumulants $U_L(\theta)$ plotted versus $\theta$
for several lattice sizes. In the vicinity of the critical point,
cumulants show a strong dependence on the system size. However, at
the critical point the cumulants adopt a nontrivial value $U^*$;
irrespective of system sizes in the scaling limit. Thus, plotting
$U_L(\theta)$ for different linear dimensions yields an
intersection point $U^*$, which gives an accurate estimation of
the critical density in the infinite system. Fig.~\ref{figure1}
shows this procedure for square lattices. The value obtained for
the critical density was $\theta_c=0.502(1)$. In addition, the
fixed value of the cumulants, $U^* =0.615(5)$, is consistent with
the extremely precise transfer matrix calculation of $U^*
=0.6106901(5)$~\cite{KAMIE} for the 2D Ising model. This finding
may be taken as a first indication of universality. However, as
recently pointed out by Selke et al.~\cite{SELKE1,SELKE2}, the
value of the cumulant intersection may depend on various details
of the model, which do not affect the universality class, in
particular, the boundary condition, the shape of the lattice, and
the anisotropy of the interactions. Consequently, more research is
required to determine the universality class of a phase
transition.


Once we know $\theta_c$, the critical exponent $\nu$ can be
calculated from the full data collapsing of $U_L$. The results are
shown in Fig.~\ref{figure2}(a), where an excellent fit was
obtained for $\nu =1$. In the inset, the data are plotted over a
wider range of temperatures, exhibiting the typical behavior of
the cumulants in presence of a continuous phase transition.

Given $\theta_c=0.502(1)$ and $\nu=1$, $\beta$ and $\gamma$ were
obtained from the collapse of the curves of $\delta$ and $\chi$,
as it is shown in Fig.~\ref{figure2}(b). The data scaled extremely
well using $\beta=1/8$ and $\gamma=7/4$. The results in
Figs.~\ref{figure1} and~\ref{figure2} support the hypothesis of 2D
Ising universality class, which is consistent with the two
competing ordered states near the transition.

We now analyze the results corresponding to triangular lattices.
In this case, there are three competing ordered states (the order
parameter has three components) and this transition is expected to
be in the universality class of the two-dimensional Potts model
with $q=3$. The inset in Fig.~\ref{figure3}(a) shows Binder
cumulants plotted versus $\theta$. The intersection point
converges to a fixed point, allowing an accurate estimation of the
critical density [$\theta_c=0.530(1)$] and the fixed value of the
cumulants [$U^* =0.613(5)$]. This value of $U^*$ is practically
indistinguishable from previous estimates for the three-state
Potts model (see for instance Ref.~\cite{TOME}, where $U^* \cong
0.613$).

On the other hand, the critical exponents were obtained from the
scaling plots of $U_L$ [Fig.~\ref{figure3}(a)], $\delta$
[Fig.~\ref{figure3}(b)] and $\chi$ [inset in
Fig.~\ref{figure3}(b)]. Very good collapses were obtained using
the exact 2D three-state Potts model scaling parameters
($\nu=5/6$, $\beta=1/9$ and $\gamma=13/9$). Our findings are
consistent with the hypothesis that this phase transition,
occurring on triangular lattices, belongs to the universality
class of the two-dimensional Potts model with $q=3$.

In summary, we have used Monte Carlo simulations and finite-size
scaling theory to resolve the universality class of the I-N phase
transition occurring in a system of long rods on square lattices
with two allowed orientations and triangular lattices with three
allowed orientations. As it was evident from the calculation of
the critical exponents and the behavior of Binder cumulants, the
universality class was shown to be that of the 2D Ising model for
square lattices and the 2-D Potts model with $q=3$ for triangular
lattices.

\acknowledgments We thank D. Dhar for his comments on an earlier
version of this paper. This work was supported in part by CONICET
(Argentina) under project PIP 6294; Universidad Nacional de San
Luis (Argentina) under project 322000 and the National Agency of
Scientific and Technological Promotion (Argentina) under project
33328 PICT 2005.

\end{document}